\documentstyle[preprint,tighten,prd,aps,eqsecnum,epsfig]{revtex}

\def\Tr{\text{Tr}\,}
\def\cD{{\cal D}}
\def\cM{{\cal M}}
\def\tr{\text{tr}\,}
\def\Str{\text{Str}\,}
\def\Sdet{\text{Sdet}\,}
\def\Seff{S_{\text{eff}}}
\def\sgn{\text{sgn}\,}

\begin{document}
\draft
\preprint{MIT-CTP-3142}
\title{Effective Lagrangian for strongly coupled 
domain wall fermions}
\author{Federico Berruto\thanks{E-mail: fberruto@bu.edu}
and Richard C. Brower\thanks{E-mail: brower@bu.edu}}
\address{Physics Department, Boston University,
590 Commonwealth Avenue, Boston, MA 02215}
\author{Benjamin Svetitsky\thanks{On leave from the
School of Physics and Astronomy, Raymond and Beverly Sackler
Faculty of Exact Sciences, Tel Aviv University, 69978 Tel Aviv, Israel.
E-mail: bqs@julian.tau.ac.il}}
\address{Center for Theoretical Physics,
Laboratory for Nuclear Science and Department of Physics,
Massachusetts Institute of Technology,
Cambridge, MA 02139}
\date{\today}
\maketitle
\begin{abstract}
We derive the effective Lagrangian for mesons in lattice gauge theory
with domain-wall fermions in the strong-coupling and large-$N_c$ limits.
We use the formalism of supergroups to deal with the
Pauli-Villars fields, needed to regulate the contributions
of the heavy fermions.
We consider the spectrum of pseudo-Goldstone bosons and show that 
domain wall fermions are doubled and massive in this regime.
Since we take the extent and lattice spacing of the fifth dimension to infinity
and zero respectively, our conclusions apply also to overlap fermions.
\end{abstract}

%
\narrowtext
\section{Introduction}
From its inception, lattice gauge theory was hampered by the absence of
an adequate fermion formulation.
It was thought for many years that
there is no discretization scheme that displays explicit 
chiral symmetry without either sacrificing locality or introducing
non-physical doublers in the fermion spectrum. 
This state of affairs was summarized in the famous no-go theorem of
Nielsen and Ninomiya~\cite{nielsen}. 
Recent years have seen major progress
in the construction of lattice regularizations of fermions with
good chiral properties~\cite{chiralreview}.
Domain wall~\cite{kaplan} and 
overlap~\cite{neuberger} formulations preserve chiral symmetry exactly,
allowing the construction of vector gauge theories on the lattice.
An idea common to these schemes is that of
employing a large number of fermion flavors, one of which survives in the
continuum limit as a massless fermion supporting exact chiral symmetry. 

In the domain wall formulation, the expanded flavor space may be seen as an
extra dimension with a defect in a background field.
One chirality of the Dirac spinor is
exponentially localized along the defect.
If the extra dimension is periodic and of finite
extent $L_5$, the other chirality will be localized along an unavoidable
second defect, while all the other Dirac fermions, $L_5-1$ in number, stay heavy. 
This is also the scheme in Shamir's surface-fermion
variant~\cite{Shamir,furmanshamir}, which we study in this paper; here the
chiral modes are localized on the surfaces of a five-dimensional slab.
In either case, the limit $L_5\rightarrow \infty$ requires that
the contribution of the heavy flavors
be subtracted~\cite{neuberger,furmanshamir,neuberger3,vranas} by introducing
$L_5$ flavors of pseudofermion fields, often called Pauli-Villars (PV) fields.
These are bosons that have the same index structure as
the fermions.

Domain wall fermions are equivalent to overlap fermions~\cite{neuberger}
in the limit
that the size and lattice spacing of the fifth dimension are taken to
infinity and zero, respectively~\cite{neuberger3,noguchi}.
The bulk degrees of freedom remain massive.
They decouple as the four-dimensional lattice is removed, 
leaving a four-dimensional Dirac operator describing the 
chiral surface modes. 
   
In this paper we study domain wall fermions in the limit of strong gauge
coupling.
We adapt the classic method of Kawamoto and Smit~\cite{kawamotosmit} to
derive an effective action for mesonic degrees of freedom.\footnote{
See also \cite{kluberg-stern1}.
This method has been elaborated and applied many times.
See for instance \cite{martin,rebbi,aoki,ichinose}.}
A novel ingredient of the domain wall formalism is the PV
bosons.
We combine the fermions and pseudofermions into a supervector, and similarly
group the effective degrees of freedom into a supermatrix.
Then the formalism of supergroups allows us easily to extend the treatment
of the fermions to an analysis of the complete theory.
The effective action contains all the meson-like states that can be constructed
from pairs of fermions, from pairs of bosons, and from boson-fermion pairs.

The interesting question is, does this theory really behave as one would
expect of a theory of exactly chiral quarks?
To address this, we take the number of colors $N_c$ to infinity and look for
saddle points of the effective action.
Following Kawamoto and Smit, we begin by setting the Wilson parameter $r$
to zero and classifying the symmetries of the theory; we display explicitly
the Goldstone bosons that are the fluctuations around the saddle point.
The spectrum of Goldstone bosons is characteristic of a theory with
complete fermion doubling, which is not surprising since it is the term
proportional to $r$ in the fermion action that breaks naive doubling in the
first place.
When we proceed, however, to perturb the saddle point with the $r$ terms,
we find that the doubling persists and that the only effect of the perturbation
is identical to that of adding a mass term
({\em \`a la\/} Shamir~\cite{Shamir,furmanshamir})
to the domain wall action.
We conclude that in the strong coupling limit, domain wall fermions
are doubled and massive, with all that that implies for the mesonic
spectrum.

We thus confirm the conclusion reached by two of us~\cite{browersvetitsky}
through study of the Hamiltonian of domain wall fermions at strong coupling.
Unfortunately, our analysis here is more involved and perhaps less
conclusive.
Study of the Hamiltonian via Rayleigh-Schr\"odinger perturbation theory is
straightforward and leads unambiguously to an effective Hamiltonian for
mesonic degrees of freedom.
The symmetries of this effective theory are readily apparent.
In the Euclidean
formulation some additional approximation is necessary, such as a hopping
parameter expansion, mean field theory, or the large-$N_c$ limit.\footnote{
In the Hamiltonian formalism, a strong coupling expansion combined 
with a large-$N_c$ expansion has been used in~\cite{smit,berruto}.}
Even at large $N_c$, one might argue that there exists another saddle point
with different symmetry properties.
Nonetheless, the Euclidean study complements the Hamiltonian
results in that it does not require a time-continuum limit and thus does
not introduce a new question regarding the order of limits.\footnote{
It is curious but true that one may consistently ignore the PV bosons
in the Hamiltonian calculation to lowest order in $1/g^2$
\cite{browersvetitsky}, but not in the Euclidean calculation.
To our knowledge, there is no established correspondence between the
perturbation series in the two formalisms.}

In Sec.~\ref{Sec2} we derive the strong-coupling effective action.
For pedagogic reasons, we begin by ignoring the PV bosons.
Following Kawamoto and Smit~\cite{kawamotosmit}, we integrate out the gauge
field in this limit
and rewrite the resulting path integral in terms of a bosonic matrix field.
That this path integral generates the same Green functions as the original
Grassmann integral is due to an equality between generating functions that was
proven in \cite{kawamotosmit} (following \cite{creutz}).
This path integral contains the effective action.
While its derivation does not depend on a large-$N_c$ limit, we give its
explicit form for this case only.
In the last part of Sec.~\ref{Sec2}
we repeat the derivation with the bosons included.
This requires a generalization to supermatrices
of the identity relating generating functions, which we prove in 
Appendix \ref{AppB}.

Since the chiral symmetry of the domain wall action
arises dynamically, it is difficult to analyze the symmetry of the effective
action directly.
The symmetry analysis cannot be divorced from the dynamics.
Section~\ref{Sec3} is devoted to deriving the symmetry breaking 
pattern of the theory and the spectrum of Goldstone bosons.
We can only make analytic progress in the $N_c\to\infty$ limit.
Here the task at hand is to find saddle points of the effective action.
Again we start with an analysis of the theory without the PV bosons.
We begin by setting $r$ to zero which makes finding a saddle point
straightforward.
In this limit the symmetry of the effective action is apparent; upon
introducing an {\it ansatz\/} for the
saddle point we display the pattern of spontaneous symmetry breaking and
the concomitant Goldstone bosons.
Then we turn on the Wilson term in the action and we 
compute its effect on the saddle point.
It turns out that the degeneracy of the $r=0$ saddle point is
broken by the $r$ term, but only in the way that a mass term breaks it,
that is, by breaking chiral symmetry while leaving the doubling.
We show explicitly that inclusion of the Shamir mass term does the same.

The extension of the calculation to include the PV bosons is straightforward.
The saddle point in the enlarged field space is the obvious extension of
the saddle point found above.
Perturbing the saddle point with the Wilson term of the bosonic fields 
gives a mass matrix for the pseudo-Goldstone bosons that is identical
in form to that due to the massive fermions.

We conclude with some discussion and comparison to other work.

\section{The Effective Action at Strong Coupling and Large $N_c$}
\label{Sec2}
\subsection{Without Pauli-Villars bosons}

We define domain wall fermions on a five-dimensional 
hypercubic lattice.
The fifth dimension has finite extent $L_5$; for generality we assign a lattice
spacing $a$ to the four 
dimensions of Euclidean space-time and a separate lattice spacing $a_5$ to
the fifth dimension.
We shall denote by
$x=(x_1,x_2,x_3,x_4)$ a coordinate vector in the four dimensional
space, with Greek indices $\mu,\nu,\ldots$;
the fifth coordinate is $s$. 
The fermion field $\psi^{a\alpha}_{xs}$ is a Grassmann field carrying
color index $a=1,\ldots,N_c$ and Dirac-flavor index $\alpha=1,\ldots,4N_f$.
The gauge field $U_{ x\mu}$ is an $SU(N_c)$ matrix field that
resides on the four-dimensional links only and is independent of $s$.

The action is a sum of pure-gauge and Dirac actions,
\begin{equation}
S(U,\bar{\psi},\psi)=S_U(U)+S_F(\bar{\psi},\psi,U).
\label{globalaction}
\end{equation}
The gauge action is as usual a sum over plaquettes,
\begin{equation}
S_U=\frac{2N_c}{g^2}\sum_P\left(1-\frac{1}{N_c}\text{Re}\,\Tr U_P\right).
\label{wilsonaction}
\end{equation}
We write the fermion action as
\begin{equation}
S_F=S_4+S_5+S_m,
\label{fermionicaction}
\end{equation} 
with\footnote{Our Dirac matrices $\gamma_\mu$
are hermitian and satisfy $\{\gamma_\mu,\gamma_\nu\}=2\delta_{\mu\nu}$.
We define $\alpha_\mu=\gamma_4\gamma_\mu$ so that $\alpha_4=1$ and
$\alpha_{1,2,3}$ are anti-hermitian. $\gamma_5$ is hermitian.
(See Appendix~\ref{AppA}.)}
\begin{eqnarray}
S_4&=&\sum_{xs\mu}\left[ \bar{\psi}^{a\alpha}_{xs}
\left(\frac{r+\gamma_\mu}{2}\right)_{\alpha\beta}U_{ x\mu}^{ab}
\psi^{b\beta}_{x+\hat\mu,s}
+\bar{\psi}^{a\alpha}_{x+\hat\mu,s}
\left(\frac{r-\gamma_\mu}{2}\right)_{\alpha\beta}U_{ x\mu}^{\dag ab}
\psi^{b\beta}_{xs}\right], 
\label{4fermionaction}\\
S_5&=&\rho \sum_x\sum_{s=1}^{L_5-1}
\left[ \bar{\psi}^{a\alpha}_{xs}
\left(\frac{1+\gamma_5}{2}\right)_{\alpha\beta}\psi^{a\beta}_{x,s+1}
+\bar{\psi}^{a\alpha}_{x,s+1}
\left(\frac{1-\gamma_5}{2}\right)_{\alpha\beta}\psi^{a\beta}_{xs}\right],
\label{5fermionaction}\\
S_m&=&(m-4-\rho)\sum_{xs}\bar{\psi}^{a\alpha}_{xs}
\psi^{a\alpha}_{xs}, \label{Mfermionaction}
\end{eqnarray}
and $\rho=a/a_5$.
Only $S_4$ contains the gauge field.

In the strong-coupling limit we drop $S_U$.
The effective action
will then be written in terms of the  meson field,
\begin{equation}
\cM^{\alpha \beta}_{ss'}(x)=\frac{1}{N_c}\sum_{a=1}^{N_c}
\psi_{xs}^{a\alpha}\bar{\psi}_{xs'}^{a\beta} \; .
\label{mesonfield}
\end{equation} 
This is a color singlet, local in $x$ but nonlocal in $s$, and hence we write
it as a $(4N_fL_5)\times(4N_fL_5)$ matrix.
(This is in line with the view that $s$ represents an internal flavor index.
Eventually we will isolate the projection of $\cM$ onto the subspace
spanned by the chiral surface modes.)

\begin{figure}[c]
\begin{center}
\vskip 1cm
\mbox{\epsfig{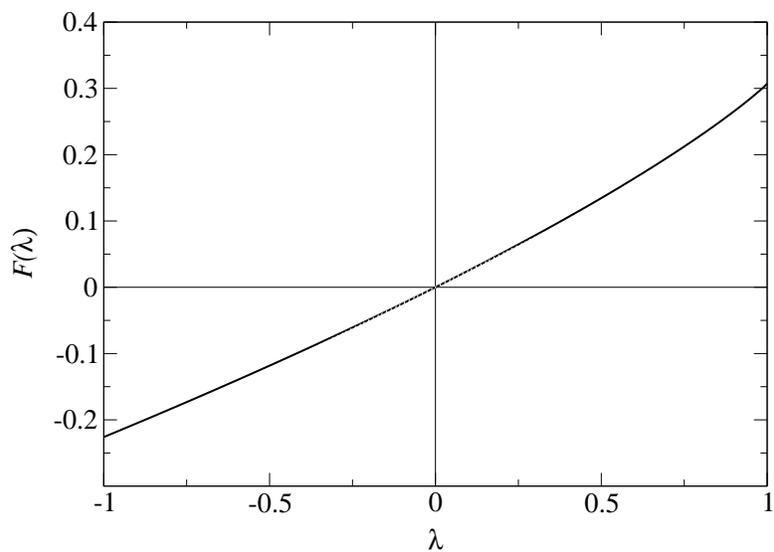}}
\vskip 1cm
\caption{The function $F(\lambda)$.  The dotted line is $F=\lambda/4$.}
\label{fig1}
\end{center}
\end{figure}

In the strong-coupling limit the partition function 
is given by
\begin{equation}
Z=\int \cD\bar{\psi}\,\cD\psi\,\cD U\, \exp S_F(\bar{\psi},\psi,U)
\label{scpartitionfunction}
\end{equation}
The integral over the gauge field can be performed immediately by noting
that it is simply a product of decoupled link integrals,
\begin{eqnarray}
e^{S_1(\bar\psi,\psi)}&\equiv&
\int \cD U\,\exp S_4(\bar{\psi},\psi,U)\nonumber\\
&=&\prod_{ x\mu}\int dU\,
\exp \Tr[\bar{A}_{\mu}(x)U+U^{\dagger}A_{\mu}(x)],
\label{ewaa}
\end{eqnarray}
where we have defined
\begin{eqnarray}
\bar{A}^{ba}_{\mu}(x)&=&\sum_s \bar{\psi}^{a\alpha}_{xs}
\left(\frac{r+\gamma_{\mu}}{2}\right)_{\alpha\beta}
\psi^{b\beta}_{x+\hat{\mu},s},\label{ba}\\
A^{ba}_{\mu}(x)&=&\sum_s\bar{\psi}^{a\alpha}_{x+\hat{\mu},s}
\left(\frac{r-\gamma_{\mu}}{2}\right)_{\alpha\beta}
\psi^{b\beta}_{xs}.\label{a}
\end{eqnarray}
The integral in Eq.~(\ref{ewaa}) can be carried out
explicitly for any value of $N_c$ \cite{BTR},
and the result expressed in terms of $\cM(x)$.
We are interested in the limit $N_c\to\infty$
\cite{browernauenberg,brezingross}.
Defining
\begin{equation}
\lambda_{\mu}(x)=-4\cM(x)\frac{r+\gamma_{\mu}}{2}
\cM(x+\hat{\mu})\frac{r-\gamma_{\mu}}{2},
\label{lambda}
\end{equation}
the result is
\begin{equation}
S_1(\cM)=N_c\sum_{x\mu}\tr F(\lambda_{\mu}(x)),
\label{s1m}
\end{equation}
where
\begin{equation}
F(\lambda)=1-\sqrt{1-\lambda}+\log\left[\frac{1}{2}\left(1+\sqrt{1-\lambda}\
\right)\right].
\label{fl}
\end{equation}
The trace (``tr'') in Eq.~(\ref{s1m}) denotes a trace over 
Dirac-flavor and $s$ indices, not to be confused with the color trace (``Tr'')
appearing in Eq.~(\ref{wilsonaction}).
We plot $F(\lambda)$ in Fig.~\ref{fig1}, and note that
one may use a straight line
$F(\lambda)\simeq \lambda/4$ as a first approximation.

The remainder of $S_F$ couples fields along the fifth axis for fixed $x$.
We can rewrite this in terms of $\cal M$ as well, {\em viz.},
\begin{equation}
S^{\perp}\equiv S_5+S_M=
-N_c\sum_x\tr D^{\perp}\cM(x),
\label{dperm}
\end{equation}
where we denote by $D^{\perp}$ the fifth-dimension Wilson-Dirac operator,
\begin{equation}
D^{\perp}_{ss'}=\rho \left[\left(\frac{1+\gamma_5}{2}\right)\delta_{s,s'-1}+
\left(\frac{1-\gamma_5}{2}\right)\delta_{s,s'+1}
\right]+(m-4-\rho)\delta_{s,s'},
\label{dper}
\end{equation}
with implicit limits on $s$ and $s'$ as indicated in Eq.~(\ref{5fermionaction}).
The complete partition function is then
\begin{equation}
Z(J)=\int \cD \bar{\psi}\,\cD \psi\,\exp\left[S_1(\cM)
+S^{\perp}(\cM) + S_J(J,\cM)\right],
\label{scpartitionfunction2}
\end{equation}
where we have added an external current coupled to mesonic bilinears according to
\begin{equation}
S_J=\sum_{xss'} \bar{\psi}^{a\alpha}_{xs}
J_{ss'}^{\alpha\beta}(x)\psi_{xs'}^{a\beta}
=-N_c\sum_x\tr J(x)\cM(x)
\label{externalcurrent}
\end{equation} 
so that functional differentiation with respect to $J$ generates Green functions of the
fermion bilinear field $\cM(x)$.

As shown in \cite{kawamotosmit}, the same Green functions are generated by
the purely bosonic partition function
\begin{equation}
Z(J)=\int \cD M(x)\,\exp\left[-N_c\sum_x\tr \log M(x)\right]
\exp\left[S_1(M)+S^{\perp}(M)+S_J(J,M)\right].
\label{bosonicpartitionf}
\end{equation}
Here $M(x)\in U(4N_fL_5)$ is a bosonic, unitary matrix field and
$\cD M(x)$ indicates the invariant Haar measure at each 4d site $x$.
(We derive this result in Appendix~\ref{AppB}, and generalize it to include
Pauli-Villars fields.)
Collecting everything, we have
\begin{equation}
Z(J)=\int \cD M(x)\,e^{\Seff(M)},
\label{partitionfunctionm}
\end{equation}
with
\begin{eqnarray}
\Seff(M)&=&N_c\left[
\sum_{x,\mu}\tr F\Bigl(-M(x)(r+\gamma_\mu)M(x+\hat\mu)(r-\gamma_\mu)\Bigr)
-\sum_x\tr D^{\perp}M(x)\right.\nonumber\\
&&\qquad\left.-\sum_x\tr\log M(x)
-\sum_x\tr J(x)M(x)\right].
\label{seffm}
\end{eqnarray}
We set $J=0$ henceforth.

The fifth-dimension Wilson-Dirac operator $D^{\perp}$ is not hermitian.
It will be convenient to work with the hermitian operator
$h\equiv \gamma_4D^{\perp}$, so we substitute
$M\to M\gamma_4$, giving
\begin{eqnarray}
\Seff(M)&=&N_c\left[ \sum_{x,\mu}
\tr F\Bigl(-M(x)\gamma_4(r+\gamma_\mu)M(x+\hat\mu)\gamma_4(r-\gamma_\mu)\Bigr)
-\sum_x\tr hM(x)\right.\nonumber\\
&&\qquad\left.-\sum_x\tr\log M(x)\right].
\label{seffm2}
\end{eqnarray}
The measure and $\tr\log M$ are unchanged since $\gamma_4$ is a
special unitary matrix. $h$ is precisely the single-site Hamiltonian
studied (in the $a_5\to0$ limit) in \cite{browersvetitsky}.  We review
its spectrum and eigenfunctions in Appendix~\ref{AppA}.

\subsection{Addition of Pauli-Villars bosons}

We have so far neglected the Pauli-Villars fields.
Introduction of supermatrix notation will enable us easily to extend our calculation
to include their effects in the effective action.

Following Vranas \cite{vranas}, we introduce scalar fields
$\phi_{xs}^{a\alpha}$ and their conjugates,
with the same indices as the fermions.
Their action $S_{PV}(\bar\phi,\phi)$ is identical in form to the fermion action
$S_F$ [see Eqs.~(\ref{fermionicaction})--(\ref{Mfermionaction})].
The only difference is the imposition of
antiperiodic boundary conditions instead of the free surfaces at
$s=1$~and~$s=L_5$, which
we write as a modification of $S_5$,
\begin{equation}
S_5\to \rho \sum_x\sum_{s=1}^{L_5}
\left[ \bar{\phi}^{a\alpha}_{xs}
\left(\frac{1+\gamma_5}{2}\right)_{\alpha\beta}\phi^{a\beta}_{x,s+1}
+\bar{\phi}^{a\alpha}_{x,s+1}
\left(\frac{1-\gamma_5}{2}\right)_{\alpha\beta}\phi^{a\beta}_{xs}\right],
\label{S5PV}
\end{equation}
where
\begin{equation}
\phi_{x,L_5+1}^{a\alpha}=-\phi_{x1}^{a\alpha}. 
\end{equation}

We group $\psi$ and $\phi$ into a superfield,
\begin{equation}
{\renewcommand{\arraystretch}{1.5}
\Psi_{xs}^{a\alpha}=\left(\begin{array}{c}
\psi_{xs}^{a\alpha} \\
\phi_{xs}^{a\alpha} \end{array}\right).
}
\label{superfield}
\end{equation}
The generalized meson field is a supermatrix containing Grassmann elements
as well as ordinary numbers,
\begin{equation}
\cM^{\alpha \beta}_{ss'}(x)=\frac{1}{N_c}\sum_{a=1}^{N_c}
\Psi_{xs}^{a\alpha}\bar{\Psi}_{xs'}^{a\beta}=
\frac{1}{N_c}\sum_{a=1}^{N_c}\left(
{\renewcommand{\arraystretch}{1.5}
\begin{array}{cc}
\psi_{xs}^{a\alpha}\bar{\psi}_{xs'}^{a\beta}& 
\psi_{xs}^{a\alpha}\bar{\phi}_{xs'}^{a\beta}\\
\phi_{xs}^{a\alpha}\bar{\psi}_{xs'}^{a\beta}& 
\phi_{xs}^{a\alpha}\bar{\phi}_{xs'}^{a\beta}
\end{array}
}
\right).
\label{mesonfieldpv}
\end{equation} 
It is a $(4N_f2L_5)\times(4N_f2L_5)$ matrix that contains the old meson
matrix~(\ref{mesonfield}) in the upper left-hand corner.

In the strong coupling limit the partition function is now
\begin{equation}
Z=\int {\cal D}\bar\psi\,{\cal D}\psi\,{\cal D}\bar\phi\,{\cal D}\phi\,
{\cal D}U\,\exp\left[
S_F(\bar\psi,\psi,U)+S_{PV}(\bar\phi,\phi,U) \right].
\label{scpartitionfunctionpv}
\end{equation}
We integrate out the gauge fields as we did in Eq.~(\ref{ewaa}),
\begin{eqnarray}
e^{S_1(\bar\psi,\psi,\bar\phi,\phi)}&\equiv&
\int \cD U\,\exp [S_4(\bar{\psi},\psi,U)+S_4(\bar\phi,\phi,U)]\nonumber\\
&=&\prod_{ x\mu}\int dU\,
\exp \Tr[\bar{A}_{\mu}(x)U+U^{\dagger}A_{\mu}(x)],
\label{ewaapv}
\end{eqnarray}
where we have defined, in analogy with Eqs.~(\ref{ba})~and~(\ref{a}),
\begin{eqnarray}
\bar{A}^{ba}_{\mu}(x)&=&\sum_s \bar{\Psi}^{a\alpha}_{xs}
\left(\frac{r+\gamma_{\mu}}{2}\right)_{\alpha\beta}
\Psi^{b\beta}_{x+\hat{\mu},s},\label{bapv}\\
A^{ba}_{\mu}(x)&=&\sum_s \bar{\Psi}^{a\alpha}_{x+\hat{\mu},s}
\left(\frac{r-\gamma_{\mu}}{2}\right)_{\alpha\beta}
\Psi^{b\beta}_{xs}.\label{apv}
\end{eqnarray}
Performing the integral (\ref{ewaapv}) in the limit $N_c\to\infty$ we obtain
[cf.~Eq.~(\ref{s1m})]
\begin{equation}
S_1(\cM)=N_c\sum_{x\mu}\Str F(\lambda_{\mu}(x)),
\label{s1mpv}
\end{equation}    
where Str denotes the supertrace (see Appendix~\ref{AppB}).
$F$ and $\lambda$ are the same as before, with $\lambda$ defined in terms of
the new supermatrix field $\cM$.
The equivalent of Eq.~(\ref{dperm}) is
\begin{equation}
S^{\perp}=
-N_c\sum_x \Str D^{\perp}\cM(x),
\label{dpermpv}
\end{equation}
where the new $D^{\perp}$ is defined by
\begin{equation}
D^{\perp}=\left(\begin{array}{cc}
 D^{\perp}_F & 0 \\
0 & D^{\perp}_{PV} \\
\end{array}\right).
\label{dpercomp}
\end{equation}
Here $D^{\perp}_F$ is the fermionic fifth-dimension
Wilson-Dirac operator given in Eq.~(\ref{dper});
the 
Wilson-Dirac operator $D^{\perp}_{PV}$ for the bosons is derived from 
Eq.~(\ref{S5PV}) and differs from $D^{\perp}_F$ only on the boundaries.
Thus we arrive again at Eq.~(\ref{scpartitionfunction2}),
\begin{equation}
Z=\int \cD \bar{\Psi}\,\cD \Psi\,\exp\left[S_1(\cM)
+S^{\perp}(\cM) \right],
\label{scpartitionfunction2S}
\end{equation}
{\em mutatis mutandis}.

We show in Appendix~\ref{AppB} that the integral over the superfield $\Psi$ can
be replaced by an integral over a supermatrix field $M$,
\begin{equation}
Z =\int \cD M(x)\,\exp\left[-N_c\sum_x\Str \log M(x)\right]
\exp\left[S_1(M)+S^{\perp}(M) \right].
\label{bosonicpartitionfpv}
\end{equation}
In this equation $M(x)$ is an element of the
supergroup $U(4N_fL_5|4N_fL_5)$ and $\cD M(x)$ is the corresponding
invariant Haar measure at each 4d site $x$.
Since $D^{\perp}$ is not hermitian, we again substitute
$M\rightarrow M\gamma_4$ and define the
hermitian operator $h=\gamma_4D^{\perp}$.
The result is
\begin{eqnarray}
\Seff(M)&=&N_c\left[ \sum_{x,\mu}
\Str F\Bigl(-M(x)\gamma_4(r+\gamma_\mu)M(x+\hat\mu)\gamma_4(r-\gamma_\mu)\Bigr)
-\sum_x\Str hM(x)\right.\nonumber\\
&&\qquad\left.-\sum_x\Str\log M(x)\right].
\label{seffm2pv}
\end{eqnarray}
This is the effective action for all the meson-like degrees of freedom.

\section{Chiral symmetry breaking pattern and Goldstone bosons}
\label{Sec3}

The large-$N_c$ limit allows us to evaluate the partition function by finding
saddle points of the effective actions (\ref{seffm2})~and~(\ref{seffm2pv}).
We shall begin by setting $r=0$, whereupon Wilson
fermions reduce to naive fermions, and then we will see how things change
as the Wilson term is restored. Again, we leave the inclusion of PV bosons to the end.

\subsection{The $r=0$ case; no PV bosons}
\label{Sec31} 

For $r=0$ the effective action (\ref{seffm2}) becomes 
\begin{equation}
\Seff(M)=N_c\left[ \sum_{x,\mu}
\tr F\Bigl(M(x)\alpha_\mu M(x+\hat\mu)\alpha_\mu\Bigr)
-\sum_x\tr hM(x)-\sum_{x}\tr\log M(x)\right].
\label{nrpartitionfunctionm2}
\end{equation}
We can eliminate the $\alpha$ matrices by
performing the following unitary transformation (spin 
diagonalization~\cite{smit,SDQW})
\begin{equation} 
M(x)\to (\alpha_1)^{x_1}(\alpha_2)^{x_2}(\alpha_3)^{x_3}
M(x)(\alpha_3)^{x_3}(\alpha_2)^{x_2}(\alpha_1)^{x_1}.
\label{unitarytransf}
\end{equation} 
We obtain the simplified expression,
\begin{equation}
\Seff(M)=N_c\left[ \sum_{x,\mu}
\tr F\Bigl(M(x)M(x+\hat\mu)\Bigr)
-\sum_x\tr hM(x)-\sum_{x}\tr\log M(x)\right].
\label{nreffectiveaction}
\end{equation}

If we set $h=0$,
the theory described by the effective action (\ref{nreffectiveaction})
is invariant under the transformation
\begin{equation}
M(x)\to\left\{
\begin{array}{ll}
UM(x)V^\dag&\quad\text{for $x$ even}\\
VM(x)U^\dag&\quad\text{for $x$ odd}
\end{array}\right.
\label{emt}
\end{equation}
where $U,V\in U(4N_fL_5)$. 
This is a global left--right symmetry typical of non-linear sigma models.
Inclusion of $h$ restricts the
transformation by the condition $Uh=hV$. Within a degenerate
eigensubspace of $h$, this means that $U=V$. 
Within the nullspace of $h$, the restriction vanishes and the symmetry under
$U\not=V$ rotations returns.
In the limit $L_5\to\infty$, $h$ possesses $4N_f$ zero modes, which are just the
chiral surface modes for the zero-dimensional domain wall.  The
symmetry realized among these modes is $U(4N_f)\times U(4N_f)$.

To translate this back into the
original fermionic coordinates we must note
that we have redefined 
$M(x)$ through left multiplication by a factor of $\gamma_4$
so that the standard chiral $U(N_f)\times U(N_f)$ transformation is
\begin{eqnarray}
\psi_{xs} &\to& e^{i(\theta_V^a\lambda^a
+\theta_A^a\lambda^a\gamma_5)}\psi_{xs}
\label{vap} ,\\
\bar{\psi}_{xs}\gamma_4 &\to& \bar{\psi}_{xs} \gamma_4
e^{-i(\theta_V^a\lambda^a +\theta_A^a\lambda^a\gamma_5)},
\label{vabp}
\end{eqnarray} 
where the $\lambda^a$ are $U(N_f)$ flavor generators.
This operation does {\em not\/} leave $h$ invariant.
The zero eigenvectors of $h$, however, are eigenvectors of $\gamma_5$ (see
Appendix \ref{AppA}) and thus the nullspace of $h$ is invariant under this
$U(N_f)\times U(N_f)$ group.
This is a subgroup of the $U(4N_f)\times U(4N_f)$ symmetry group of the
nullspace, in fact, of its $U(4N_f)$ subgroup defined by $U=V$.

The group $U(4N_f)\times U(4N_f)$
is the symmetry of fully doubled, naive fermions.
Evidently for
$r=0$ the domain wall formalism yields ordinary doubling.

\begin{figure}[c]
\begin{center}
\vskip 1cm
\mbox{\epsfig{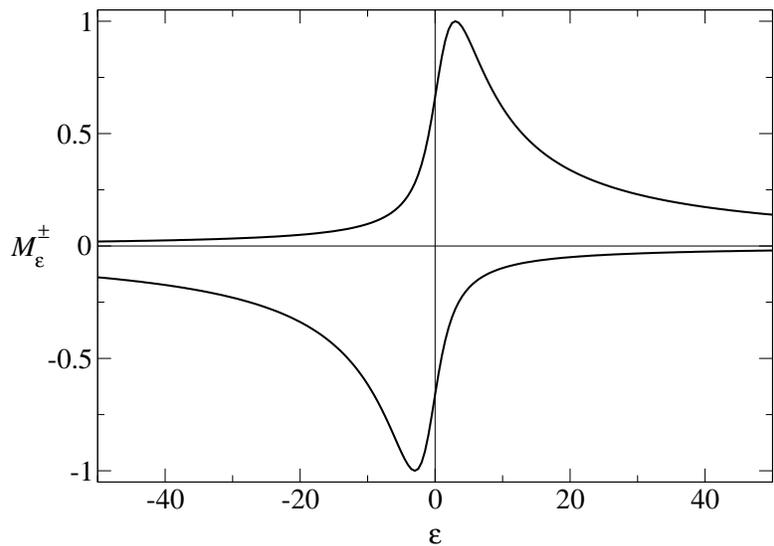}}
\vskip 1cm
\caption{The mean field solutions $M_\epsilon^+$ (upper curve)
and $M_\epsilon^-$ (lower curve).}
\label{fig2}
\end{center}
\end{figure}

These symmetry considerations guide us in choosing an {\em ansatz\/} for the
saddle point of $\Seff$.
Let us consider the eigenstate basis of the single-site Hamiltonian $h$,
\begin{equation}
h|\epsilon\rangle=\epsilon |\epsilon\rangle.
\label{heigenspace}
\end{equation}
We suppose that, at the saddle point, $M(x)$ is translation-invariant and
diagonal in the $h$ basis,
\begin{equation}
M(x)=\hat M=\sum_\epsilon|\epsilon\rangle M_\epsilon\langle \epsilon|.
\label{ansatzm}
\end{equation}
The effective action (\ref{nreffectiveaction}) now reads
\begin{equation}
\Seff(M)=N_cV \sum_\epsilon \left[ dF(M_\epsilon^2)-\log M_\epsilon
  -\epsilon M_\epsilon \right],
\label{aeffectiveaction}
\end{equation}
where $d=4$ is the number of dimensions.
Using the formula (\ref{fl}) for $F$, we differentiate
with respect to $M_\epsilon$ to find the roots
\begin{equation}
M_\epsilon^{\pm}=\frac{\epsilon(d-1)\pm
  \sqrt{\epsilon^2(d-1)^2+(2d-1)(d^2+\epsilon^2)}}{d^2+\epsilon^2} .
\label{stationarypoints}
\end{equation}
In the subspace of zero modes of $h$, where $\epsilon=0$,
this reduces to
\begin{equation}
M_0^{\pm}=\pm \frac{\sqrt{2d-1}}{d}.
\label{mzm}
\end{equation} 
We plot $M^\pm(\epsilon)$ in Fig.~\ref{fig2}, and note that $|M^\pm|<1$ for
any $\epsilon$.
To maximize the action at the saddle point, we choose
\begin{equation}
M_\epsilon=\left\{
\begin{array}{ll}
M_\epsilon^+&\quad\text{for $\epsilon>0$}\\
M_\epsilon^-&\quad\text{for $\epsilon<0$}.
\end{array}\right.
\label{choice}
\end{equation}
This choice satisfies
\begin{equation}
M_{-\epsilon}=-M_\epsilon.
\label{MmM}
\end{equation}

The solution (\ref{stationarypoints})--(\ref{choice})
does not break any $U=V$ symmetries [see Eq.~(\ref{emt})] within degenerate
eigensubspaces of $h$, but it does break the $U\not=V$ symmetries in the
space of zero modes according to $U(4N_f)\times U(4N_f)\to U(4N_f)$.
We may use a $U\not=V$ transformation to set $M_0=M_0^+$ without loss of
generality, and so $\hat M=M_0{\bf1}$ within the space of zero modes.

In order to display the Goldstone bosons, we allow $M(x)$ to fluctuate about
the saddle point $\hat M$.
We write
\begin{equation}
M(x)=\left\{
\begin{array}{ll}
\hat M\Sigma(x)&\quad\text{for $x$ even}\\
\hat M\Sigma^\dag(x)&\quad\text{for $x$ odd}
\end{array}\right.
\label{smallosc1}
\end{equation}
and
\begin{equation}
\Sigma(x)=e^{iH(x)}.
\label{smallosc2}
\end{equation}
We take the Goldstone field $H(x)$ to be non-zero only within the zero modes
of $h$, and thus it belongs to the algebra $U(4N_f)$; it commutes with $\hat M$.
Inserting Eqs.~(\ref{smallosc1}) and~(\ref{smallosc2}) into the action
(\ref{nreffectiveaction}), we expand to
second order in $H(x)$ to obtain\footnote{The term linear in $H$ that comes
of expanding $F$ cancels against the logarithmic term
in Eq.~(\ref{nreffectiveaction}).
In $d=4$ the coefficient in square brackets in Eq.~(\ref{Ssmallosc}) is
equal to $31/192$.}
\begin{equation}
\Seff(M)\simeq\Seff(\hat M)-\frac12\left[F'(M_0^2)M_0^2+F''(M_0^2)M_0^4\right]
\sum_{x\mu}\tr[H(x)-H(x+\hat\mu)]^2.
\label{Ssmallosc}
\end{equation}
This is a quadratic action representing massless Goldstone bosons in the
adjoint representation of $U(4N_f)$.

If the field $H$ is taken to represent fluctuations {\em outside\/}
the $U(4N_f)$
subgroup, it acquires a mass of the order of $m-4$, indicating that these bosons
are not Goldstone bosons.
We may sum up the results so far by stating that domain wall fermions with
$r=0$ possess the symmetry of naive fermions and yield Goldstone bosons
according to the simplest scheme of spontaneous symmetry breaking.

\subsection{The $r\ne 0$ case; no PV bosons}
\label{Sec32}
 
Turning on the Wilson term in the action will break the
$U(4N_f)\times U(4N_f)$ symmetry.
In principle, we should redo the above analysis and find a new saddle point
of the complete action.
To simplify the calculation, however, we will stay within the manifold of saddle
points of the $r=0$ action, and determine the effective action that selects
the point of lowest energy in this manifold.
The symmetry of this effective action will be the symmetry left unbroken
by the $r$ terms.

We begin again with the action (\ref{seffm2}), this time keeping $r\not=0$.
After spin diagonalization (\ref{unitarytransf}),
the strong-coupling effective action takes the form
[cf.~Eq.~(\ref{nreffectiveaction})]
\begin{eqnarray}
\Seff(M)&=&N_c\left[ \sum_{x\atop{k=1,2,3}}
\tr F\Bigl(M(x)[1-(-1)^{x_k}r\gamma_k]M(x+\hat k)[1-(-1)^{x_k}r\gamma_k]\Bigr)
\right. \nonumber\\
&&\qquad+\sum_x
\tr F\Bigl(M(x)[1+s_3(x)r\gamma_4]M(x+\hat4)[1-s_3(x)r\gamma_4]\Bigr) \nonumber\\
&&\qquad\left.-\sum_{x\atop{\ }}\tr hM(x)-\sum_x\tr\log M(x)\right]
\label{nreffectiveaction_r}\\
&=&\Seff^{r=0}(M)+\Delta\Seff(M).
\end{eqnarray}
where $s_3(x)\equiv(-1)^{x_1+x_2+x_3}$.
For the sake of simplicity we shall approximate $F(\lambda)\simeq \lambda/4$,
which gives us
\begin{eqnarray}
\Delta\Seff(M)&=&\frac{N_c}4\left\{\sum_{x\atop{k=1,2,3}}\left[
-r(-1)^{x_k}\,\tr \gamma_k\{M(x),M(x+\hat k)\}
+r^2\,\tr M(x)\gamma_kM(x+\hat k)\gamma_k
\right]\right.\nonumber\\
&&\qquad\left.+\sum_{x\atop{\ }}\left[
-rs_3(x)\,\tr \gamma_4[M(x),M(x+\hat 4)]
-r^2\,\tr M(x)\gamma_4M(x+\hat 4)\gamma_4\right]\right\}.
\label{DSeff}
\end{eqnarray}

We stay with the {\em ansatz\/} $M(x)=\hat M$ as given by
Eqs.~(\ref{ansatzm})--(\ref{choice}), translation-invariant
and diagonal in the eigenbasis of $h$.
We leave free the orientation of $\hat M$ within the space of zero modes.
The resulting effective action is
\begin{equation}
\Delta\Seff(\hat M)=N_c\frac{r^2}4\left[
\sum_{x\atop{k=1,2,3}}\tr  \hat M\gamma_k\hat M\gamma_k
-\sum_{x}\tr \hat M\gamma_4\hat M\gamma_4\right].
\label{DSeffMhat}
\end{equation}

\begin{figure}[c]
\begin{center}
\vskip 1cm
\mbox{\epsfig{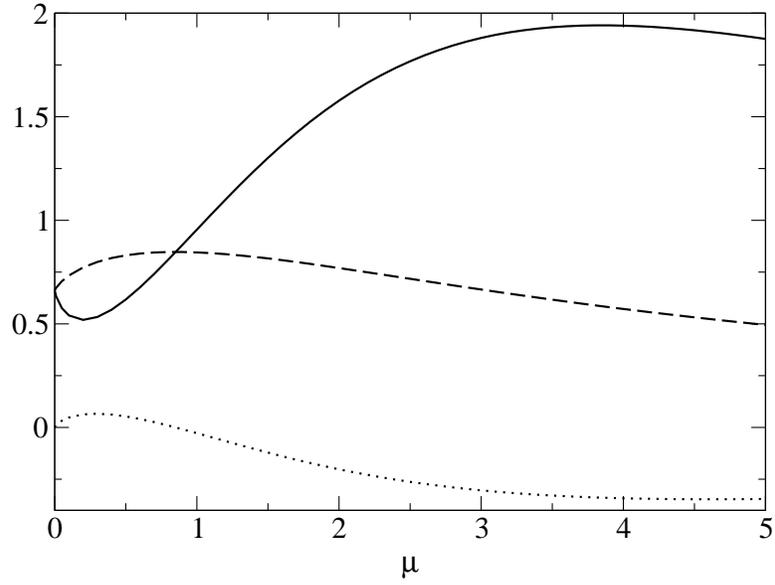}}
\vskip 1cm
\caption{The integrals ${\cal J}$ (dashed curve) and ${\cal Y}$ (dotted
curve), and the combination ${\cal J}-4{\cal Y}$ (solid curve), as functions
of $\mu=m-4$.}
\label{fig3}
\end{center}
\end{figure}

We collect in Appendix~\ref{AppA} some formulas connected with diagonalization
of $h$ (see \cite{browersvetitsky}).
We work in the limit $a_5\to0$, where $h$ becomes
\begin{equation}
h=\gamma_4\gamma_5\frac\partial{\partial s}+(m-4)\gamma_4.
\label{hcont}
\end{equation}
In the limit $L\equiv L_5a_5\to\infty$, the $4N_f$
surface states of $h$ become exact zero modes.
We denote them (for each flavor)
by $|0^\eta\sigma\rangle$, where $\eta=\pm$ is the sign of the energy before
taking $L\to\infty$, and $\sigma$ is a spin label.
We label states in the continuum spectrum of $h$ as $|\epsilon\sigma\rangle$;
they satisfy $h|\epsilon\sigma\rangle=\epsilon|\epsilon\sigma\rangle$, with
$|\epsilon|>m-4$.
It is straightforward to evaluate the traces in Eq.~(\ref{DSeffMhat})
in the basis $|0^\pm\sigma\rangle,|\epsilon\sigma\rangle$.
Using the key formula (\ref{simplif}), we find that the first term in
Eq.~(\ref{DSeffMhat}) is independent of the orientation of $\hat M$ in the
zero modes $|0^\eta\sigma\rangle$;
the second term yields
\begin{equation}
\Delta\Seff(\hat M)
= {\it const.}-N_c\frac{r^2}2\sum_x
\langle0^\eta\sigma|\gamma_4\hat M\gamma_4|0^{\eta'}\sigma'\rangle
\langle0^{\eta'}\sigma'|\hat M|0^\eta\sigma\rangle.
\label{4thtermMhat}
\end{equation}
Only the second matrix element in Eq.~(\ref{4thtermMhat}) depends on the
orientation of $\hat M$ in the zero modes.
As we show in Appendix \ref{AppA}, the coefficient matrix
$\langle0^\eta\sigma|\gamma_4\hat M\gamma_4|0^{\eta'}\sigma'\rangle$
takes the form ${\cal J}\rho_3$ in the indicated basis for the zero modes,
where ${\cal J}$ is plotted in Fig.~\ref{fig3}.

Now we change basis \cite{browersvetitsky}
in the space of zero modes from
$|0^\pm\sigma\rangle$ to 
$|0^{L,R}\sigma\rangle\equiv\frac1{\sqrt2}(|0^+\sigma\rangle\pm|0^-\sigma\rangle)$.
The new basis states are localized on either boundary and hence represent the
true chiral modes.
The change of basis converts $\rho_3$ to $\rho_1$ (which is $\gamma_4$ in the
chiral basis for the Dirac matrices).
Hence the effective action is
\begin{equation}
\Delta\Seff(\hat M)={\it const.}-N_c\frac{r^2}2{\cal J}
\sum_x\tr\rho_1 \hat M.
\label{SeffH}
\end{equation}
Alternatively, one may remove the $\rho_1$ from Eq.~(\ref{SeffH}) by
means of a $U(4N_f)\times U(4N_f)$ transformation
with $U=\rho_1$, $V=1$.
Then the new term in the action takes the form $\tr \hat M$.
In any case, it is clear that $\Delta\Seff$ breaks the $U(4N_f)\times U(4N_f)$
symmetry to $U(4N_f)$.
No chiral symmetry is left.

Another way to see the destruction of chiral symmetry is to treat
$\Delta\Seff$ as a perturbation on the symmetric action $\Seff^{r=0}$.
As we have seen, the latter possesses $(4N_f)^2$ Goldstone bosons;
the new term renders all these bosons massive.

These results show the same pattern as the symmetry breaking in a theory of
naive fermions due to a fermion mass term.
The chiral $U(4)\times U(4)$ symmetry of massless naive fermions is broken
by a mass term $m_f\bar\psi\psi$ to $U(4)$.
The latter is a vector symmetry, and one expects no Goldstone bosons.
This is our central result:
Domain wall fermions in strong coupling behave as naive, doubled fermions
with a mass term.

The same symmetry-breaking pattern emerges from addition of a Shamir
mass term \cite{Shamir} to the domain-wall action.
This couples the fields on the boundaries
$s=0,L_5$ according to the action
\begin{equation}
S_S= -\rho m_S\sum_x\left[\bar\psi_{xL_5}\left(\frac{1+\gamma_5}2\right)\psi_{x0}
+\bar\psi_{x0}\left(\frac{1-\gamma_5}2\right)\psi_{xL_5}\right].
\label{shamirmassaction}
\end{equation}
 In the $a_5\to0$ limit, this adds a
term $h_S$ to the single-site Hamiltonian $h$.  Its matrix element
between wave functions $\psi(s)$ and $\psi'(s)$ is
\begin{equation}
\langle\psi|h_S|\psi'\rangle=\bar\psi(0)\psi'(L)
+\bar\psi'(0)\psi(L),
\end{equation}
and it contributes a term to the effective action,
\begin{equation}
\Seff^S(M)=-N_c\sum_x\tr h_SM(x).
\end{equation}
Perturbing within the $m_S=0$ ground states, we write
\begin{equation}
\tr h_SM(x)=\tr h_S\hat M.
\end{equation}
and allow $\hat M$ to rotate within
the zero-mode sector as above.
As shown in \cite{browersvetitsky}, within the space of zero modes
the operator $h_S$ takes the form $m_S(m-4)\rho_1$,
as found above for the term induced at strong coupling.

\subsection{PV bosons restored}

Returning to the full effective action (\ref{seffm2pv}),
we can step quickly through our analysis above of the pure fermionic theory.
We begin with the case $r=0$.
The site--site coupling term in $\Seff$ is then symmetric under
the graded unitary group
$U(4N_fL_5|4N_fL_5)\times U(4N_fL_5|4N_fL_5)$.
The $h$ term breaks this symmetry to a vector subgroup, except within the
space of the fermionic zero modes where a $U(4N_f)\times U(4N_f)$ chiral
symmetry survives as before.
The pseudofermion modes are all massive and thus do not add any chiral symmetry
to $\Seff$.
We eliminate the $\alpha$ matrices from the action via spin diagonalization,
Eq.~(\ref{unitarytransf}).

We assume a saddle point at which the homogeneous mean field commutes with $h$,
\begin{equation}
M(x)=\hat M=\sum_\epsilon|\epsilon\rangle M_\epsilon\langle \epsilon|.
\end{equation}
Just as $h$ has two blocks, one for the fermionic modes and one for the bosons,
so does $M_\epsilon$.
Note that this diagonal supermatrix has no non-commutative elements that are
nonzero.
Since the mean field equations stemming from Eq.~(\ref{aeffectiveaction})
decouple the modes of $h$ from each other,
the solution for $M_\epsilon$ is unchanged from
Eq.~(\ref{stationarypoints}).\footnote{
There are two blocks in $M_\epsilon$, but both are given by the formula
(\ref{stationarypoints}).}
In particular, the Pauli-Villars modes have no effect on the fermionic modes.
The chiral $U(4N_f)\times U(4N_f)$ symmetry of the fermionic zero modes
is broken spontaneously to $U(4N_f)$, bringing about the appearance of
$16N_f^2$ Goldstone bosons in the adjoint representation of $U(4N_f)$.

Restoring the $r$-dependent terms in $\Seff$, we perturb the saddle
point.
The $r$-dependent terms in the effective action are [cf.~Eq.~(\ref{DSeff})]
\begin{eqnarray}
\Delta\Seff(M)&=&\frac{N_c}4\left\{\sum_{x\atop{j=1,2,3}}\left[
-r(-1)^{x_j}\,\Str \gamma_j\{M(x),M(x+\hat \jmath)\}
+r^2\,\Str M(x)\gamma_jM(x+\hat \jmath)\gamma_j
\right]\right.\nonumber\\
&&\quad\left.+\sum_{x\atop{\ }}\left[
-rs_3(x)\,\Str \gamma_4[M(x),M(x+\hat 4)]
-r^2\,\Str M(x)\gamma_4M(x+\hat 4)\gamma_4\right]\right\}.
\label{DSeffpv}
\end{eqnarray}
Again we set $M(x)=\hat M$ and allow $\hat M$ to rotate in the space
of zero modes of $h$. We evaluate the supertraces in Eq.~(\ref{DSeffpv}) in the
basis of fermion states $|0^{\pm}\sigma\rangle$,$|\epsilon\sigma\rangle_F$ and
boson states $|\epsilon k\sigma\rangle_B$.
We use Eq.~(\ref{simplif}) again in the evaluation of
Eq.~(\ref{DSeffpv}), but now
Eqs.~(\ref{simplifpv1}) and~(\ref{sgnepspv1}) introduce new, nonzero matrix
elements. 
Dependence on $\hat M$ survives in both $O(r^2)$ terms in
Eq.~(\ref{DSeffpv}), which yield 
\begin{eqnarray}
\Delta \Seff(\hat{M})&=& \frac{N_c}{2}r^2 \sum_x 
\Str \left(\sum_{j=1}^3\gamma_j \hat{M}\gamma_j \hat{M} 
-\gamma_4 \hat{M}\gamma_4 \hat{M}\right)
\nonumber\\
&=& \frac{N_c}{2}r^2 \sum_x 
\left(\sum_j\langle 0^\eta \sigma
  |\gamma_j\hat{M}\gamma_j|0^{\eta'}\sigma'\rangle
-\langle 0^\eta \sigma|\gamma_4\hat{M}\gamma_4|0^{\eta'}\sigma'\rangle\right)
\nonumber\\
&&\qquad\qquad\qquad\times
\langle 0^{\eta'} \sigma'|\hat M|0^{\eta}\sigma\rangle.
\label{Seffbosons}
\end{eqnarray}    
We evaluate the matrix elements in Appendix~\ref{AppA} to obtain the final
result
\begin{equation}
\Delta\Seff(\hat M)={\it const.}-N_c\frac{r^2}2\left( {\cal J}-4{\cal Y}\right)
\sum_x\tr\rho_1\hat M.
\end{equation}
Only the coefficient has changed from Eq.~(\ref{SeffH}).  One might
have expected that the well-known cancellation between the fermions
and the Pauli-Villars bosons affects the low-energy sector, but this
does not occur.  Nonetheless, the heavy superpartners of the mesonic
bound states must play an essential role in canceling the infinite
number of heavy modes, thus allowing the above truncation to a
low-energy effective theory.

\section{Discussion}
\label{finaldiscussion}

In this paper we have argued that, at strong coupling, non-Abelian
gauge theory with domain wall fermions exhibits explicit chiral
symmetry breaking but with an enlarged spectrum of pseudo-Goldstone modes
characteristic of the doubling phenomenon.  We
derived the effective action of domain wall fermions at the leading
order of the strong coupling and large-$N_c$ expansions in the limit
that the lattice spacing of the fifth dimension goes to zero and its extent to
infinity; therefore our conclusions apply also
to overlap fermions.  

Our results also depend on perturbing to lowest
order in the Wilson $r$ parameter in
the neighborhood of a stationary point with maximal
symmetry consistent with $r = 0$. 
Another phase, with different symmetry, might be encountered at finite $r$.
We consider it unlikely that moving to a saddle point with lower symmetry will
restore any symmetry broken at this most symmetric saddle point.
Thus it would be surprising if chiral symmetry is restored at the true 
$r\not=0$ saddle point.
On the other hand, the $U(4N_f)$ symmetry characteristic of exact doubling
could end up broken; likewise, corrections of higher order in $1/g^2$ might
break this symmetry.
We do not know of another signal of doubling that could then be sought in the
meson spectrum.

Our analytic results provide evidence for a phase transition
from a weak coupling phase with good chiral properties to a strong
coupling phase where the desired chiral properties are lost.
This is based on the observation that to
lowest order in $r$ there is an explicit breaking term
identical in form to the quark mass term introduced by Shamir. Thus
the true Goldstone modes are lost. In the Monte Carlo simulations of
domain wall fermions, a pattern of increasing violation of chiral
symmetry has been observed as the coupling is made stronger. The
violations of chiral symmetry are suppressed exponentially in
$L_5$ and therefore increasing the coupling
requires increasing $L_5$ and consequently
the computational burden. 
Our calculation, however, has been done in the
$L_5\to\infty$ limit so we should not be seeing
this problem. 
In this limit, domain wall fermions are
equivalent to overlap fermions. The locality of the (effective) 4d
Dirac operator is only guaranteed~\cite{localneuberger} at moderate
couplings and therefore the bad chiral properties at strong coupling
might be connected with the fact the theory is no longer  local.

Since the induced breaking of chiral symmetry is identical in form to
that caused by a Shamir mass term, it is tempting to consider tuning the
latter to cancel the former.
This has been considered, in the context of finite-$L_5$ effects, by
the authors of \cite{aokietal}.
One is reminded of Wilson fermions, where the pions
may be made massless by tuning the hopping parameter.
Wilson fermions do {\em not\/} regain chiral symmetry by this procedure, except
in the continuum limit;
the masslessness of the pion stems from the proximity of a phase where
parity and flavor symmetry are spontaneously broken \cite{aoki}.
If the analogy is exact, then tuning $m_S$ to make the Goldstone bosons
massless will not restore chiral symmetry in the lattice domain-wall theory.
Neither would it restore locality to the effective overlap operator.

Recently, overlap fermions have been studied with a hopping parameter
expansion, valid for large values of $m$ \cite{ichinosenagao}.
Our coefficients $\cal J$ and $\cal Y$ vanish in the limit $m\to\infty$,
so that chiral symmetry is apparently restored and our results are consistent
with those of \cite{ichinosenagao}.
Golterman and Shamir \cite{golterman}, however,
have argued that 
a chiral $U(N_f)\times U(N_f)$ symmetry remains {\em rigorously\/} unbroken
for sufficiently large (not necessarily infinite) values of $m$.
Thus there should be at least $N_f^2$ Goldstone bosons,
including a $U(1)$ pseudoscalar.
Our results, while in agreement with the earlier
Hamiltonian study \cite{browersvetitsky}, stand in contradiction with those of
\cite{golterman} in the common
region of applicability, namely, large $g$ and $m$.
We can only conjecture some subtlety in taking the double limit
$g^2,m\to\infty$.

Our strong coupling results rest on the analysis of a chiral $U(4 N_f
L_5|4 N_f L_5)$ super matrix model on a 4d lattice.
An external field is provided by the Hamiltonian in the fifth dimension.
Such matrix models are very interesting but notoriously hard to analyze.
This supermatrix model reformulation of
domain wall fermions is not peculiar to strong coupling. At any
coupling, after integrating out the gauge fields, $U(N_c)$
Yang-Mills theory with domain wall fermions must take the form of an effective
theory,
\begin{equation}
Z = \int \cD M(x) \exp\left[S_4(M) - N_c \sum_x\Str\log M(x)
- N_c \sum_x\Str hM(x)\right],
\end{equation}
expressed in terms of meson-like superfields. 
This is the ineluctible consequence of local 4d gauge
invariance.
The effective theory contains all the mesonic bound
states---$(4 N_f L_5)^2$ spin-flavors of quark-antiquark pairs,
a like number of pseudoquark-antipseudoquark pairs, and their superpartners.
The physically
interesting limit is $4 N_f L_5\to\infty$, an infinite number of
``flavors.''

Changing from $U(N_c)$ to $SU(N_c)$ will 
introduce baryon-like superfields but these are most likely
irrelevant to the low energy chiral phase structure. With this
{\em caveat\/}, the only difference between strong coupling and general
coupling is that now the 4d action, $S_4(M)$, is no longer restricted
to nearest-neighbor couplings, becoming more and more non-local as we
approach weak coupling. We expect this exact representation for $U(N_c)$ domain
wall lattice QCD to be in the correct chiral phase
at weak enough coupling.  The central
question of this paper, whether the correct chiral symmetry
pattern persists at strong coupling, is a question concerning
the role of the non-locality for a $U(n|n)$ 4d supermatrix model in
an external field at large $n$. Our present results give some insight
for the nearest-neighbor matrix model derived in strong coupling near
$r = 0$. More general analysis of these matrix models can help
to understand domain wall fermions in the confined phase.

\section*{Acknowledgements}

We thank Michael Creutz, Maarten Golterman, and Yigal Shamir for helpful
discussions.
This work was supported in part by the Department of Energy under
Contracts No.~DE-FG02-91ER4067A6 and No.~DF-FC02-94ER40818.
The work of F.~B. is supported by an INFN Postdoctoral Fellowship.
He thanks the members of the Theoretical Particle Physics
Group at Boston University for their kind hospitality.

\appendix
\section{}
\label{AppA}

We use the following Dirac matrices for the Euclidean theory:
\begin{equation}
\begin{tabular}{*{2}{p{2.4in}}}
$\gamma_k=\rho_2\sigma_k,\quad k=1,2,3$&$\gamma_4=\rho_1$\\
$\alpha_k=\gamma_4\gamma_k=i\rho_3\sigma_k$&$\alpha_4\equiv1$\\
$\gamma_5=\rho_3$&$\sigma_{\mu\nu}=i\gamma_\mu\gamma_\nu$\\
\end{tabular}
\end{equation}
The single site fermion Hamiltonian is the same as that studied in
\cite{browersvetitsky},
\begin{equation}
h=\gamma_4\gamma_5\frac{\partial}{\partial s}+\mu\gamma_4,
\label{hsmall}
\end{equation}
where $\mu\equiv m-4$.
We impose the boundary conditions
\begin{equation}
(1+\gamma_5)u(L)=(1-\gamma_5)u(0)=0.
\label{bc}
\end{equation}
on the eigenfunctions of $h$.
We have taken a continuum limit $a_5\to0$, $L_5\to\infty$, with $L=a_5L_5$
held fixed.

Since $\gamma_4\gamma_5=-i\rho_2$, there is no spin dependence in $h$ and
thus its eigenfunctions take the form
\begin{equation}
u=\left(\begin{array}{c}\hat f(s)\chi\\\hat g(s)\chi\end{array}\right)
\label{spinor}
\end{equation}
where $\chi$ is any 2-spinor.
The spectrum contains discrete surface states with wave functions
\begin{eqnarray}
\hat f&=&\pm A_0\sinh\kappa(s-L)\equiv \pm f_0(s) \label{f0}\\
\hat g&=&A_0\sinh \kappa s\equiv g_0(s), \label{g0}
\end{eqnarray}
where $\kappa$ satisfies the eigenvalue condition
\begin{equation}
\kappa=\mu\tanh\kappa L.
\end{equation}
Their energies are
\begin{equation}
\epsilon=\pm\epsilon_0=\pm\frac \mu{\cosh\kappa L}.
\end{equation}
As $L\to\infty$, we have $\kappa\to \mu$ and $\epsilon_0\to0$.
Taking spin into account,
there are four zero-energy states for each flavor, which we denote
by $|0^\eta\sigma\rangle$, where $\eta=\pm$ is the sign of the energy before
taking $L\to\infty$, and $\sigma$ is a spin label.

There are also continuum modes,
\begin{eqnarray}
\hat f&=&\pm A_k\sin k(s-L)\equiv \pm f_\epsilon(s)\\
\hat g&=&A_k\sin ks\equiv g_\epsilon(s),
\end{eqnarray}
with the quantization condition
\begin{equation}
k=\mu\tan kL
\label{quant}
\end{equation}
and energies
\begin{equation}
\epsilon=\pm\frac \mu{\cos kL}=\pm\sqrt{k^2+\mu^2}.
\end{equation}
There is a gap to this pseudo-continuum, $|\epsilon|>\mu=m-4$.
When $k\gg \mu$ the solutions of Eq.~(\ref{quant}) approach $(2n+1)\pi/2L$.
We label these states $|\epsilon\sigma\rangle$.
Note that we take $k>0$ and that there are two energies $\epsilon$ for
each value of $k$.

The relation $\{\gamma_5,h\}=0$ implies that
\begin{equation}
\gamma_5|0^\eta\sigma\rangle=|0^{-\eta}\sigma\rangle\quad\text{and}\quad
\gamma_5|\epsilon\sigma\rangle=|-\epsilon,\sigma\rangle.
\label{g5}
\end{equation}
Together with orthogonality of the
eigenfunctions, this relation can also be used to prove
\begin{equation}
\int ds\,f_0f_\epsilon=\int ds\,g_0g_\epsilon=0.
\end{equation}
Moreover, the overlap integral
\begin{equation}
\int_0^L ds\,f_0(s)g_0(s)
\end{equation}
goes to zero exponentially as $L\to\infty$.
Using the explicit forms of the $\gamma$ matrices, we find
\begin{equation}
\langle0^\eta\sigma|\gamma_\mu|0^{\eta'}\sigma'\rangle=
\langle0^\eta\sigma|\gamma_k|\epsilon\sigma'\rangle=0,
\label{simplif}
\end{equation}
for $\mu=1,\ldots,4$ and $k=1,2,3$.
On the other hand,
\begin{equation}
\langle0^\eta\sigma|\gamma_4|\epsilon\sigma'\rangle=\left\{
\begin{array}{ll}
\displaystyle{2\eta\delta_{\sigma\sigma'}}{\cal I}(\epsilon)
&\quad\text{if $\sgn\epsilon=\eta$}\\
0&\quad\text{otherwise}
\end{array}\right.
\label{sgneps}
\end{equation}
where
\begin{equation}
{\cal I}(\epsilon)=\int_0^Lds\,f_0(s)g_\epsilon(s)\simeq
-\sqrt{\frac{\mu}{L}}\frac{k}{k^2+\mu^2}
\end{equation}
for large $L$.

We use these matrix elements to calculate the coefficient matrix in the
symmetry-breaking term in the effective action (\ref{4thtermMhat}).
This matrix is
\begin{equation}
\langle0^\eta\sigma|\gamma_4\hat M\gamma_4|0^{\eta'}\sigma'\rangle
=\sum_{\epsilon\tau}\langle0^\eta\sigma|\gamma_4|\epsilon\tau\rangle
M_\epsilon\langle\epsilon\tau|\gamma_4|0^{\eta'}\sigma'\rangle.
\label{mat4}
\end{equation}
According to Eq.~(\ref{sgneps}), the matrix elements of $\gamma_4$ are
zero unless $\eta=\sgn\epsilon$ and $\eta'=\sgn\epsilon$;
moreover, there is no spin structure in $\gamma_4$ or in $M_\epsilon$ and
so $\sigma=\tau=\sigma'$.
Thus the matrix is diagonal, and we need calculate only
$\langle0^\eta\sigma|\gamma_4\hat M\gamma_4|0^\eta\sigma\rangle$.
Using Eq.~(\ref{g5}), we have
\begin{eqnarray}
\langle0^{-\eta}\sigma|\gamma_4\hat M\gamma_4|0^{-\eta}\sigma\rangle
&=&\langle0^\eta\sigma|\gamma_4\gamma_5\hat M\gamma_5\gamma_4|0^\eta\sigma\rangle
\nonumber\\
&=&\sum_\epsilon
\langle0^\eta\sigma|\gamma_4\gamma_5|\epsilon\sigma\rangle
\hat M_\epsilon\langle\epsilon\sigma|\gamma_5\gamma_4|0^\eta\sigma\rangle
\nonumber\\
&=&\sum_\epsilon
\langle0^\eta\sigma|\gamma_4|-\epsilon,\sigma\rangle
\hat M_\epsilon\langle-\epsilon,\sigma|\gamma_4|0^\eta\sigma\rangle
\nonumber\\
&=&\sum_\epsilon
\langle0^\eta\sigma|\gamma_4|\epsilon,\sigma\rangle
\hat M_{-\epsilon}\langle\epsilon,\sigma|\gamma_4|0^\eta\sigma\rangle
\nonumber\\
&=&\sum_\epsilon
\langle0^\eta\sigma|\gamma_4|\epsilon,\sigma\rangle
(-\hat M_\epsilon)\langle\epsilon,\sigma|\gamma_4|0^\eta\sigma\rangle,
\nonumber
\end{eqnarray}
by virtue of Eq.~(\ref{MmM}).
Thus
\begin{equation}
\langle0^{-\eta}\sigma|\gamma_4\hat M\gamma_4|0^{-\eta}\sigma\rangle
=-\langle0^\eta\sigma|\gamma_4\hat M\gamma_4|0^\eta\sigma\rangle,
\end{equation}
and the matrix takes the form ${\cal J}\rho_3$.
Explicitly,
\begin{eqnarray}
{\cal J}&=&\langle0^+{+}|\gamma_4\hat M\gamma_4|0^++\rangle\nonumber\\
&=&\int_0^\infty
d\epsilon\,\rho(\epsilon)\langle0^+{+}|\gamma_4|\epsilon+\rangle
M_\epsilon\langle\epsilon{+}|\gamma_4|0^++\rangle\nonumber\\
&=&\frac{2\mu}\pi\int_0^\infty dk\,\frac{k^2}{(k^2+\mu^2)^2}
\frac{6\sqrt{k^2+\mu^2}+8\sqrt{k^2+\mu^2+7}}{k^2+\mu^2+16}
\end{eqnarray}
when $d=4$.
Here $\rho(\epsilon)=(L/\pi)(\epsilon/k)$ is the
density of states.

Let us turn to the Pauli-Villars bosons.
Their wave functions satisfy
Eq.~(\ref{hsmall}) with antiperiodic boundary conditions,
\begin{equation}
u(L)=-u(0).
\label{abc}
\end{equation}
The spectrum contains only continuum modes,
\begin{eqnarray}
\hat f=\pm Ae^{iks}\equiv\pm f_k(s)\\
\hat g=Ae^{iks}\equiv g_k(s),
\label{pvcm}
\end{eqnarray}
where the momenta are quantized as $k=\pm(2n+1)\pi/L$.
The energies of these modes are 
\begin{equation}
\epsilon=\pm \sqrt{k^2+\mu^2},
\end{equation}
and therefore there is a gap $|\epsilon | > \mu=m-4 $.
$k$ runs from $-\infty$ to $\infty$, and there are two energies $\epsilon$
for each value of $k$.
The Pauli-Villars fields exhibit no zero mode. 

We will need to know the overlap
integrals between the continuum Pauli-Villars modes and the fermion zero modes. 
For large $L$,
\begin{eqnarray}
\int_0^L ds\,f_0(s)g_k(s)&\simeq &
-\sqrt{\frac{\mu}{2L}}\,\frac{1}{k^{2}+\mu^2}(\mu+ik)
\label{ofpv1}\\
\int_0^L ds\, f_k(s)g_0(s) 
&\simeq & \sqrt{\frac{\mu}{2L}}\,\frac{1}{k^{2}+\mu^2}(-\mu+ik).
\label{ofpv3}
\end{eqnarray}
By using these integrals and the
explicit form of the $\gamma$ matrices we find\footnote{We mark the bosonic
continuum states with $B$ subscripts and include $k$ in the label because
of the $k\to-k$ degeneracy; an $F$ subscript is added to fermion
states where needed to avoid confusion.}
\begin{eqnarray}
\langle0^\eta\sigma|\gamma_j|\epsilon k\sigma'\rangle_B
&=& i\sigma^j_{\sigma \sigma'}\sqrt{\frac{\mu}{2L}}\,\frac{1}{k^2+\mu^2}
\left[\mu \left(\eta-\sgn \epsilon \right)+i k \left(\eta +\sgn
\epsilon\right) \right],
\label{simplifpv1} \\
\langle0^\eta\sigma|\gamma_4|\epsilon k\sigma'\rangle_B
&=&\delta_{\sigma \sigma'}\sqrt{\frac{\mu}{2L}}\,\frac{1}{k^2+\mu^2}
\left[\mu \left(-\eta-\sgn\epsilon \right)+i k \left(-\eta +\sgn
\epsilon \right) \right].
\label{sgnepspv1}
\end{eqnarray}

Proceeding to the effective action (\ref{Seffbosons}),
let us first evaluate the matrix
\begin{equation}
\langle 0^\eta \sigma |\gamma_j\hat{M}\gamma_j|0^{\eta'}\sigma'\rangle = 
\sum_{\epsilon k\tau}\langle 0^\eta \sigma |\gamma_j|\epsilon k\tau\rangle_B 
M_{\epsilon}\,_B\langle\epsilon k\tau|\gamma_j|0^{\eta'}\sigma'\rangle.
\label{matrixk}
\end{equation}
Using Eqs.~(\ref{MmM}) and~(\ref{simplifpv1})
one can easily prove that the matrix 
(\ref{matrixk}) takes the form ${\cal Y}\rho_3$.
The coefficient is
\begin{eqnarray}
{\cal Y}&=&\langle0^+{+}|\gamma_j\hat M\gamma_j|0^++\rangle\nonumber\\
&=&\int_{-\infty}^\infty
d\epsilon \,\rho(\epsilon){\textstyle\frac12}\sum_{k=\pm}
\langle0^+{+}|\gamma_j|\epsilon k+\rangle_B
M_{\epsilon}\,_B\langle\epsilon k{+}|\gamma_j|0^++\rangle\nonumber\\
&=&\frac{2\mu}\pi\int_0^\infty dk\,\frac{k^2-\mu^2}{(k^2+\mu^2)^2}
\frac{6\sqrt{k^2+\mu^2}+8\sqrt{k^2+\mu^2+7}}{k^2+\mu^2+16}.
\end{eqnarray}
It is very similar to ${\cal J}$, and it is independent of the index $j$.

We can now evaluate the matrix
\begin{eqnarray}
\langle 0^\eta \sigma |\gamma_4\hat{M}\gamma_4|0^{\eta'}\sigma'\rangle &=& 
\sum_{\epsilon \tau}\langle 0^\eta \sigma |\gamma_4|\epsilon \tau\rangle_F 
M_{\epsilon}\,_F\langle\epsilon \tau|\gamma_4|0^{\eta'}\sigma'\rangle\nonumber
\\
&&+\sum_{\epsilon k\tau}\langle 0^\eta k\sigma |\gamma_4|\epsilon \tau\rangle_B 
M_{\epsilon}\,_B\langle\epsilon k\tau| \gamma_4|0^{\eta'}\sigma'\rangle .
\label{matrix4}
\end{eqnarray}
The first sum in Eq.~(\ref{matrix4}) is the same as that in
Eq.~(\ref{mat4}) and is equal to ${\cal J}\rho_3$. 
Using Eqs.~(\ref{MmM}) and~(\ref{sgnepspv1}) we find that the second sum 
equals $-{\cal Y}\rho_3$.
Changing basis in the space of zero modes as in Eq.~(\ref{SeffH})
converts $\rho_3\rightarrow 
\rho_1$.

\section{}
\label{AppB}

In this appendix we derive the basic identity for the generating function
that allows the fermionic Grassmann integrals to be replaced by
integrals over a unitary matrix field
$M(x)$ [see  Eqs.~(\ref{scpartitionfunction2}) and~(\ref{bosonicpartitionf})].
In particular we present the generalization of this
well-known result to supermatrices \cite{superstuff}, which we require
for the inclusion of the Pauli-Villars fields
[see Eqs.~(\ref{scpartitionfunction2S}) and~(\ref{bosonicpartitionfpv})].

In the absence of Pauli-Villars fields, we seek the ``replacement''
of the fermion bilinears with matrices $M(x)\in U(4N_fL_5)$,
\begin{equation}
\cM^{\alpha \beta}_{ss'}(x)=\frac{1}{N_c}\sum_{a=1}^{N_c}
\psi_{xs}^{a\alpha}\bar{\psi}_{xs'}^{a\beta} \longrightarrow M(x),
\end{equation} 
The essential identity allows the replacement of the generating function 
\begin{equation}
Z(J)=\int d\bar{\psi}\,d\psi\,\exp\left(-N_c\,\tr  J \cM\right)
=(\det J)^{N_c}
\end{equation} 
by the  $U(n)$ group integral
\begin{equation}
I(J)=\int dM\,\frac{1}{(\det M)^{N_c}} \,\exp\left(-N_c\,\tr JM\right),
\label{bosontheorem}
\end{equation}
up to a multiplicative constant,
where $dM$ is the Haar measure for $U(n)$ and $n = 4 N_f L_5$.
This identity was proven by Kawamoto and Smit\cite{kawamotosmit}
and it follows immediately from a similar identity for the one-link
integral given earlier by Creutz\cite{creutz}.

The generating function that arises in the theory with Pauli-Villars fields is
\begin{equation}
Z(J)=\int d\bar{\psi}\,d\psi\,d\bar{\phi}\,d\phi\,\exp\left(-N_c\,\Str  J \cM\right)
=(\Sdet J)^{N_c},
\end{equation}
where ${\cal M}$ is the supermatrix given in Eq.~(\ref{mesonfieldpv}).
The generalization to supermatrices of Eq.~(\ref{bosontheorem}) is 
\begin{equation}
I(J)=\int dM\,\frac{1}{(\Sdet M)^{N_c}}\exp\left(-N_c\,\Str JM\right),
\end{equation}
where $dM$ is the invariant Haar measure for the supergroup $U(n|n)$.
The latter is the group that leaves invariant the norm
$\sum_i z^*_i z_i + \sum_i
\theta^*_i \theta_i$ in complex superspace, where $\theta_i$ are Grassmann
variables and $z_i$ are c-numbers.

We choose definitions of the supertrace and superdeterminant that
differ from the conventional choice by $\Str \rightarrow - \Str$ and
$\Sdet \rightarrow 1/\Sdet$ in order to simplify the
comparison between expressions before and after the including the Pauli-Villars
fields.\footnote{ See Efetov~\cite{superstuff}.}
To be explicit, we write
\begin{equation}
M =\left(\begin{array}{cc}
A&B\\
C&D\\
\end{array}\right),
\end{equation}
with the supervector index in the order
$(\theta_1, \theta_2 \cdots \theta_n,z_1,z_2 \cdots z_n)$, {\em i.e.}, Grassmann
variables first.
Then
\begin{equation}
\Str M \equiv \tr A-\tr D,
\label{Str}
\end{equation}
and
\begin{equation}
\Sdet M\equiv e^{\Str\log M}=\frac{\det A}{\det(D - C A^{-1} B)}.
\label{Sdet}
\end{equation}

The proof of  our identity relies on the following observations.
Since the integral measure is compact and the integrand is
free of singularities, the integral $I(J)$ is holomorphic in 
the components of $J$.
In particular, if we write
\begin{equation}
J =\left(
{\renewcommand{\arraystretch}{1.5}
\begin{array}{cc}
J_{\psi \bar{\psi}}& J_{\psi \bar{\phi}}\\
J_{\phi \bar{\psi}}& J_{\phi \bar{\phi}}
\end{array}
}
\right)
\end{equation}
in block form and replace $J_{\psi \bar{\psi}} \rightarrow
z J_{\psi \bar{\psi}}$ and $J_{\phi \bar{\phi}} \rightarrow
z^{-1} J_{\phi \bar{\phi}}$,
we can expand $I(J,z)$ in a Laurent series,
\begin{equation}
I(J,z)  = \sum^\infty_{m= - \infty}c_m z^m,
\end{equation}
where $I(J) \equiv I(J,z=1)$.
There can be no dependence on $z^*$.

Next we factor the Haar measure 
for $ U(n|n) \equiv U(1) \times U(n|n)/U(1)$ as
$dM = d\phi\,d\hat M$.
The explicit parameterization for this
$U(1)$ factor is 
\begin{equation}
M =\left(\begin{array}{cc}
e^{i\phi/2} \hat A&\hat B\\
\hat C&e^{-i\phi/2}\hat D\\
\end{array}\right).
\end{equation}
where  $\hat M = M(\phi = 0)$ 
is special unitary, $\Sdet\hat M = 1$.
Note that $\Sdet M = e^{i n \phi}\,\Sdet\hat M =e^{i n \phi}$.

Introducing the source  rotated under $U(1)$,
\begin{equation}
\hat J =\left(
{\renewcommand{\arraystretch}{1.5}
\begin{array}{cc}
z e^{i\phi/2}J_{\psi \bar{\psi}}& J_{\psi \bar{\phi}}\\
J_{\phi \bar{\psi}}& z^{-1} e^{-i\phi/2}J_{\phi \bar{\phi}}
\end{array}
}
\right),
\end{equation}
the group integral becomes
\begin{equation}
I(J,z)  = \int d\phi\,d\hat M\,\frac{1}{e^{iN_c n \phi}}
\exp\left(-N_c\,\Str\hat J \hat M\right) =
\int d\phi\,e^{-iN_c n \phi}F(\hat J).
\end{equation}
Moreover, invariance of the integral $d\hat M$ over $SU(n|n)=U(n|n)/U(1)$
implies that
the kernel $F(\hat J)$ is invariant under right and left rotations,
\begin{equation}
F(\hat J) = F(U \hat J V^\dag),
\end{equation}
with $U, V \in SU(n|n)$.
Thus $F(\hat J)$ must be a function of
the invariants $\Sdet\hat J$ and $\Str(\hat J \hat J^\dag)^l$. 
Holomorphy rules out the trace terms $\Str(\hat J \hat J^\dag)^l$
because of their explicit dependence on  $z^*$.
Then using $\Sdet\hat J = ( z e^{i\phi})^n\,\Sdet J$,
we see that the integral over $\phi$ projects out a single term in
the Laurent expansion, $c_m$ for $m = N_c n$.
Hence
\begin{equation}
I(J,z=1)  = c_m(\Sdet J)^{N_c},
\end{equation}
proving the identity.

In the remainder of this appendix we shall prove
Eqs.~(\ref{s1m})~and~(\ref{s1mpv}), which 
are of paramount importance in the derivation of the effective actions 
(\ref{seffm2})~and~(\ref{seffm2pv}).
The proof of Eq.~(\ref{s1m}) was
given in~\cite{kawamotosmit} and we shall briefly sketch it here 
in order to extend it to the case of Eq.~(\ref{s1mpv}),
in which one has also to deal with the Pauli-Villars fields.
Using Eq.~(\ref{ewaa}), we express
the action (\ref{s1m}) as a sum over one-link integrals,
\begin{equation}
S_1(\bar{\psi},\psi)=\sum_{x,\mu}w\left(\bar{A}(x),A(x)\right)=\sum_{x,\mu}\log \int dU\,
\exp \Tr\left[\bar{A}_{\mu}(x)U+U^{\dagger}A_{\mu}(x)\right],
\label{s1app}
\end{equation}
where $\bar{A}$ and $A$ were defined in Eqs.~(\ref{ba})~and~(\ref{a}). 
The one-link integral appearing in Eq.~(\ref{s1app}) was computed in
the large-$N_c$ limit in~\cite{browernauenberg,brezingross}, with the result
\begin{equation}
w(\bar{A},A)=N_c^2\left\{{\textstyle\frac34}-c+{2}{N_c}\sum_{a}(c+x_a)^{1/2}
-\frac{1}{2N_c^2}\sum_{a,b}\log
\left[(c+x_a)^{1/2}+(c+x_b)^{1/2}\right]\right\}.
\label{wapp}
\end{equation}
Here $x_a$ are the eigenvalues of $\bar{A}A/N_c^2$ and $c$ is defined
implicitly by
\begin{equation}
1=\frac{1}{2N_c}\sum_a(c+x_a)^{-1/2}.
\label{capp}
\end{equation}
By expanding in $x_a$, Eqs.~(\ref{wapp})~and~(\ref{capp}) can be rewritten as a
series in 
\begin{equation}
\sum_a(x_a)^k=\Tr\left(\frac{\bar{A}A}{N_c^2}\right)^k.
\label{xatr}
\end{equation}
By rearranging the fermion fields in $\bar{A}$ and $A$ we obtain
\begin{equation}
\Tr\left(\frac{\bar{A}A}{N_c^2}\right)^k=
-\tr\left[-{\textstyle\frac14}\lambda(x) \right]^k,
\label{trtrapp}
\end{equation}
where $\lambda(x)$ is defined in Eq.~(\ref{lambda}).
Thus for any function that admits a series expansion we can write
\begin{equation}
\frac{1}{N_c}\sum_af(x_a)=f(0)-\frac{1}{N_c}
\tr\left[f\left(-{\textstyle\frac14}\lambda\right)-f(0)\right],
\label{expansionf}
\end{equation} 
and using Eqs.~(\ref{wapp})--(\ref{expansionf}) we obtain Eq.~(\ref{s1m}).

By the same method we can prove
Eq.~(\ref{s1mpv}) in the presence of the Pauli-Villars fields. The proof
relies on the fact that
Eqs.~(\ref{wapp})--(\ref{expansionf}) still hold but the
trace tr should be replaced by the supertrace Str.
Moreover $\bar{A}$ and $A$
are now defined by Eqs.~(\ref{bapv})~and~(\ref{apv}), and ${\cal M}$ by
Eq.~(\ref{mesonfieldpv}).
One can easily verify that in presence of Pauli-Villars fields one has,
instead of Eq.~(\ref{trtrapp}),
\begin{equation}
\Tr\left(\frac{\bar{A}A}{N_c^2}\right)^k=
-\Str\left[-{\textstyle\frac14}\lambda(x) \right]^k.
\label{trtrapp2}
\end{equation}
Equation~(\ref{trtrapp2}) allows the expansion of Eq.~(\ref{wapp}) in
powers of $x_a$ in the supermatrix case as well.
One can thus prove that the action can be completely
expressed in terms of $\lambda$ as in Eq.~(\ref{s1mpv}), where now $\lambda$
is a bilinear function of the supermatrix ${\cal M}$ given in
Eq.~(\ref{mesonfieldpv}).

\end{document}